\documentclass[prl,aps,reprint,graphicx,longbibliography]{revtex4-1}

\usepackage[utf8]{inputenc}
\usepackage{amsmath}
\usepackage{amssymb}
\usepackage{physics}
\usepackage{gensymb}
\usepackage{times}
\usepackage{graphicx,float,wrapfig}
\graphicspath{{img/}}
\usepackage{soul}
\usepackage{nameref}

\AtBeginDocument{%
    \newwrite\bibnotes
    \def\bibnotesext{Notes.bib}
    \immediate\openout\bibnotes=\jobname\bibnotesext
    \immediate\write\bibnotes{@CONTROL{REVTEX41Control}}
    \immediate\write\bibnotes{@CONTROL{%
    apsrev41Control,author="08",editor="1",pages="1",title="0",year="1"}}
     \if@filesw
     \immediate\write\@auxout{\string\citation{apsrev41Control}}%
    \fi
}%

\pdfoutput=1

\usepackage{xcolor}

\usepackage[colorinlistoftodos]{todonotes}
\usepackage[colorlinks=true, allcolors=blue]{hyperref}

\begin{document}

\title{Entanglement-based 3D magnetic gradiometry with an ultracold atomic scattering halo}
\author{D.~K.~Shin$^{1}$, J.~A.~Ross$^{1}$, B.~M.~Henson$^{1}$, S.~S.~Hodgman$^{1}$, and A.~G.~Truscott$^{1}$}
\email{andrew.truscott@anu.edu.au}
\affiliation{\normalsize{$^{1}$Research School of Physics and Engineering, Australian National University, Canberra 0200, Australia}}
\date{\footnotesize \today}

\begin{abstract}
Ultracold collisions of Bose-Einstein condensates can be used to generate a large number of counter-propagating pairs of entangled atoms, which collectively form a thin spherical shell in momentum space, called a scattering halo.
Here we generate a scattering halo composed almost entirely of pairs in a symmetric entangled state and observe a scattering angle-dependent mixing to the anti-symmetric state due to the presence of an inhomogeneous magnetic field.
We report on a proof-of-principle application of the observed parity dynamics to demonstrate magnetic gradiometry, insensitive to common-mode fluctuations of the background magnetic field.
Furthermore, the highly multimode nature and narrow radial width of the scattering halos enable 3D tomography of an interrogated field without the need for a scanning probe.
\end{abstract}
\maketitle  

Quantum correlations such as entanglement or squeezing can enable measurement device sensitivities that outperform the standard quantum limit (SQL) of classically correlated systems \cite{Giovannetti2004, Pezze2018}, and even the realisations of classically-forbidden tasks \cite{Horodecki2009}.
In squeezing, the improvement in sensing is due to the suppression of quantum fluctuations of the measurement variable below that of a classical state at the cost of amplified uncertainty of the complementary property \cite{Walls1983}.
Striking nonclassical features such as nonlocality exist in other forms of strongly entangled systems such as the Bell singlet $\ket{\Psi^-} = \frac{1}{\sqrt{2}} \left( \ket{\uparrow\downarrow} - \ket{\downarrow\uparrow} \right)$, which is central to quantum technologies like quantum computing and cryptography \cite{Horodecki2009}.
The strong quantum correlations can enable extremely sensitive measurements \cite{Kotler2014}, and furthermore are known to be able to achieve the fundamental limit of precision known as the Heisenberg-limit \cite{Giovannetti2004}.

Diverse areas in physics harness such correlations for quantum-enhanced sensing of gravitational waves \cite{ligo2013}, time \cite{Louchet-Chauvet2010}, and electromagnetic fields \cite{Wasilewski2010}, for example.
Among these, quantum-assisted magnetometry is an active area with tested platforms including superconducting circuits \cite{Danilin2018}, nuclear magnetic resonance \cite{Jones2009}, nitrogen-vacancy centres in diamond \cite{Steinert2010}, optomechanical microcavities \cite{Li2018}, trapped ions \cite{Ruster2017}, atomic vapours \cite{Wasilewski2010}, and ultracold atoms \cite{Ockeloen2013,Muessel2014}.
Excellent wide-field measurements of magnetic field have been investigated in nitrogen-vacancy centres in diamond \cite{Steinert2010} and ultracold atomic systems \cite{Yang2017}. 
Such magnetic microscopes show promising applications in medical and materials science, where a precise mapping of the magnetic field, and microscopic spatial resolution are simultaneously desired \cite{Meltzer2017}.
Ultracold atom microscopes rely on reconstruction of the magnetic field via imaging density modulations in elongated trapped ensembles \cite{Wildermuth2005,Yang2017}, or in-trap atom interferometry \cite{Ockeloen2013} for AC magnetometry, whilst scanning the trapped cloud (or sample relative to atoms) over the interrogation area.
So far, however, demonstrations of wide field-of-view magnetic imaging has been limited to 2D, where quantum resources have not been exploited.
A 3D magnetic field microscope may allow new applications over the current state-of-the-art 2D devices, which for tackling inverse problems of the Biot-Savart law are limited to reconstructions of 2D current distributions \cite{Roth1989}.

Here we report on proof-of-principle demonstration of entanglement-based magnetic gradiometry using strongly entangled pairs of atoms created from a collision between two Bose-Einstein condensates (BECs) \cite{Shin2018}.
The particular Bell entanglement of the pairs scattered in the collision allows an intrinsically differential measurement of the field. 
In addition, we achieve a 3D tomography of the magnetic field gradient (and absolute field strength by Ramsey interferometry) with microscopic precision of $\approx (35~\mathrm{\mu m})^3$, based on the free-expansion dynamics of the scattering halo.

\begin{figure}[!htbp]
    \centering
    \includegraphics[width=0.8\linewidth]{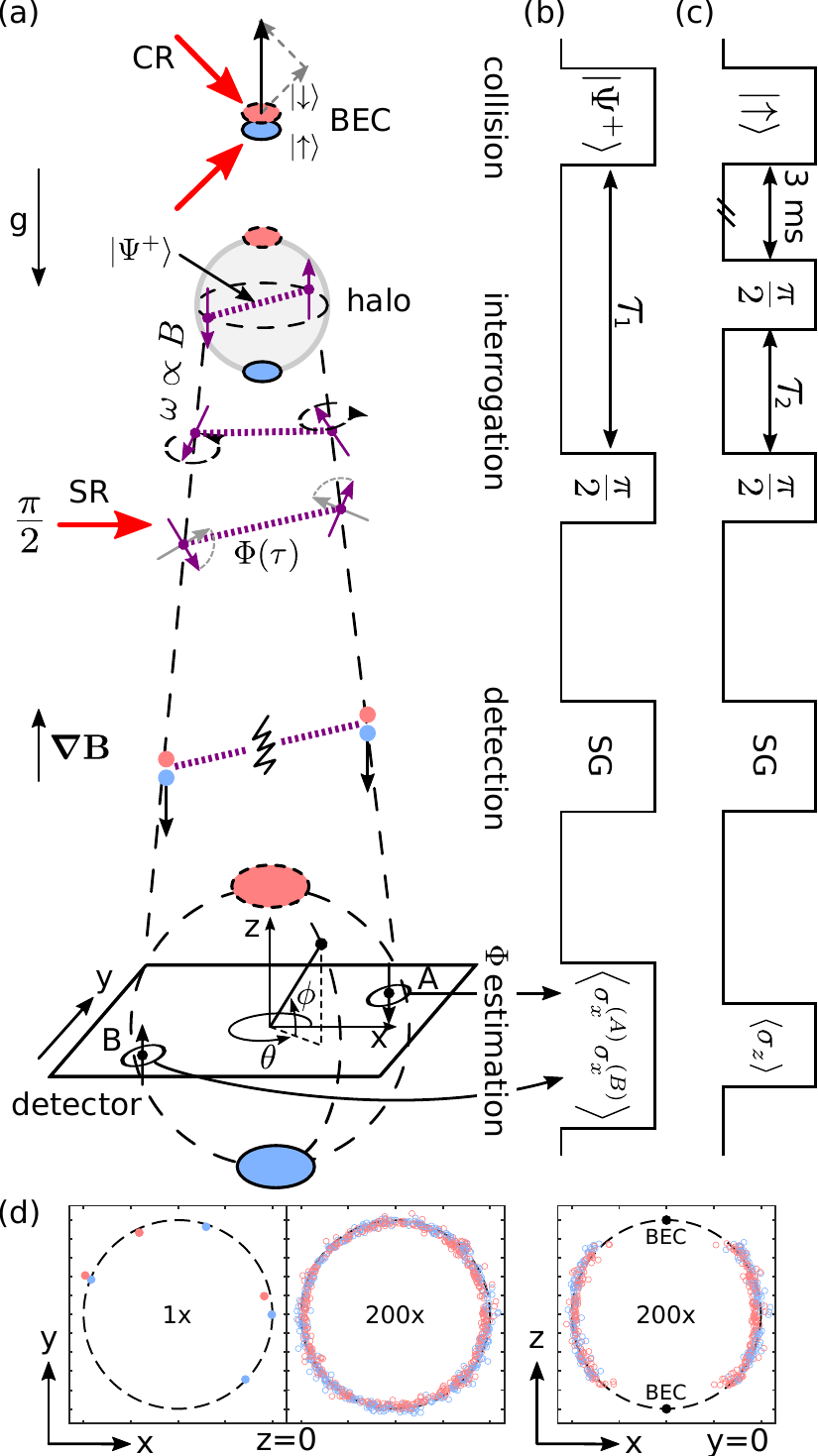}
    \caption{
    (a) Experimental schematic for entanglement-based 3D magnetic gradiometry. 
    A pair of Raman beams (CR, red arrows) induce the collision of BECs between $\ket{\uparrow}$ (blue) and $\ket{\downarrow}$ (red) atoms, forming a scattering halo composed of entangled pairs of atoms (purple arrows) in the spin triplet state $\ket{\Psi^+}$.
    Each pair accrues a phase based on its local magnetic field, such that a relative phase $\Phi$ develops across the pairs as they move through an inhomogeneous field.
    At the end of interrogation $\tau$, every atoms' spin is rotated by a $\pi/2$-pulse by the SR beams (horizontal red arrow), followed by a SG sequence to spatially separate the spin-components.
    A single-atom sensitive detector, located in the far-field, resolves individual atom's 3D position and spin.
    Regions on the scattering halo are labelled by spherical polar angles $\theta$ (azimuthal) and $\phi$ (elevation).
    Spin correlations across the pair (A, B) are evaluated to estimate $\Phi$.
    (b) Experimental sequence for (a). 
    (c) Experimental sequence to run an alternate magnetometry sequence based on Ramsey interferometry.
    (d) Planar slices of single-atom reconstructed scattering halo from a single and 200 stacked shots.
    The planar slices are averaged over a transverse width $\pm 5\%$ of the halo's radius, and atoms near the BECs ($|\phi|>50^{\circ}$) are excluded due to detector saturation.
    See main text for details.}
    \label{fig:exp_schematic}
\end{figure}

Our experiment starts with a BEC of $^4\!$He, magnetically trapped in the $m_J=+1$ sublevel of the $2\,^3\mathrm{S}_1$ metastable ground state (He$^*$) \cite{Dall2007}. 
A uniform DC magnetic field of $\mathrm{B}(\mathbf{r}) \approx 0.5~\textrm{G}$ is stabilised inside the vacuum chamber by three sets of actively controlled Helmholtz coils, which attenuate stray ac background fields over $100$-fold, and stabilise shot-to-shot variations to less than 0.1~mG \cite{Dedman2007}.
We use a two-photon stimulated Raman transition with two beams crossed at $90^{\circ}$ (CR: collision Raman) to split the BEC into two halves counter-propagating at velocities $\pm v_r \approx 60~\textrm{mm/s}$ along $\hat{z}$ in the centre of mass frame (CM) (see Fig.~\ref{fig:exp_schematic}(a)).
All our Raman transitions couple the $m_J=0$ and $m_J=+1$ sublevels \cite{Shin2018}, which are hereafter denoted by $\ket{\downarrow}$ and $\ket{\uparrow}$, respectively.
As the two BECs collide during their separation, pairs of atoms spontaneously scatter via $s$-wave collisions to form a uniformly-distributed thin spherical shell in momentum space termed a scattering halo \cite{Perrin2007}.
We set the CR beam polarisation and detuning to realise a collision between (1) $\ket{\uparrow}$-polarised BECs to prepare a $\ket{\uparrow}$-polarised coherent spin halo (Fig.~\ref{fig:exp_schematic}(c)), and (2) anti-polarised BECs ($\ket{\uparrow}$ and $\ket{\downarrow}$) to generate $\ket{\Psi^+} = \frac{1}{\sqrt{2}} \left( \ket{\uparrow\downarrow} + \ket{\downarrow\uparrow} \right)$ entangled halo (Fig.~\ref{fig:exp_schematic}(b)) as in \cite{Shin2018}.

We interrogate the magnetic field by a sequence of spin rotation pulses on the freely-expanding scattering halo.
We implement the spin rotations on the individual atoms by a second two-photon Raman transition using co-propagating beams (SR: spin Raman), which leaves the momentum unchanged (see Fig.~\ref{fig:exp_schematic}(a)), as in our previous work \cite{Shin2018}.
The SR beams are collimated with a $1/\mathrm{e}^2$ intensity radius of $\approx 2.3~\textrm{mm}$, ensuring uniform intensity and phase over the entire atomic ensemble ($\lesssim 0.4~\textrm{mm}$), to achieve a uniform spin rotation for every atom in the halo \cite{Shin2018}. 

Different pulse sequences are implemented for the two schemes reported here, namely entanglement-based magnetic gradiometry (Fig.~\ref{fig:exp_schematic}(b)) and magnetometry (Fig.~\ref{fig:exp_schematic}(c)).
Each sequence is detailed below, but first we describe our detection procedure.

After the interrogation, a Stern-Gerlach (SG) sequence separates the different $m_J$-sublevels and the 3D position of each atom is detected in the far-field (416~ms free-fall) with a quantum efficiency of $\eta \approx 0.1$, where each $m_J$-scattering halo is a thin spherical shell (see Fig.~\ref{fig:exp_schematic}(d)).
We obtain the spin-position resolved atom number distribution $n_{\uparrow/\downarrow}(\mathbf{r})$ using a conical integration volume (apex at the centre of the halo and axis along $\mathbf{r}$), such that the radial resolution is given by the thickness of the scattering halo ($\Delta r/r \approx 0.03$ \cite{SOMs}).
The angular size of the integration volume (half-cone angle $\alpha$) is matched to the spatial uncertainty expected from free-expansion of the halo \cite{SOMs}.
We find that the spatial resolution at the interrogation point is limited by the width of the collision source to $\sigma_{\mathrm{BEC}}/\sqrt{2} \approx35~\mathrm{\mu m}$, while the field-of-view corresponds to the spatial extent of the scattering halo itself, a spherical shell expanding at $2v_r\approx120~\textrm{mm/s}$ \cite{SOMs}.

The interrogation sequence for magnetometry begins $3~\mathrm{ms}$ after the $\ket{\uparrow}$-halo is created -- at which point the halo diameter has expanded to be $D \approx 360~\mu$m -- when we apply a $\pi/2$ rotation pulse which prepares each atom in an equal superposition of $\ket{\uparrow}$ and $\ket{\downarrow}$ (see Fig.~\ref{fig:exp_schematic}(c)).
A relative phase $\phi$ accumulates between $\ket{\uparrow}$ and $\ket{\downarrow}$ at the local Larmor frequency $\omega = \gamma \mathrm{B}$, where $\gamma \approx 2.8~\textrm{MHz/G}$ is the gyromagnetic ratio of He$^*$. 
A second $\pi/2$-pulse, applied after a variable delay $\tau$, encodes the interferometric phase in the normalised polarisation 
$P = (n_{\uparrow}-n_{\downarrow})/(n_{\uparrow}+n_{\downarrow})$ 
by 
\begin{equation}
    P = C \cos \phi = C \cos \gamma \mathrm{B} \tau, 
    \label{eqn:ramsey_p_vs_t}
\end{equation}
where $C$ is the interferometric contrast.

On the other hand, the entangled pair $\ket{\Psi}$ oscillates coherently between the Bell states $\ket{\Psi^+}$ (symmetric) and $\ket{\Psi^-}$ (anti-symmetric) at the difference in Larmor frequencies of each atom, such that any uniform fluctuations to the magnetic field does not affect the pair's dynamics, but rather only the difference in magnetic field experienced by each atom in the pair (i.e. the gradient across the halo along the scattering axis) \cite{Kielpinski2001,Ruster2017}.
After time $t$ has elapsed following the collision, the pair is then given by
\begin{equation}
    \ket{\Psi(t)} = \cos\Phi(t) \ket{\Psi^+} + i \sin\Phi(t) \ket{\Psi^-},
    \label{eqn:bell_state_t_evo}
\end{equation}
where the Bell mixing angle evolves according to $\Phi(t) = \gamma/2 \int_{0}^{t} \delta\mathrm{B}(\tau) d\tau$, and $\delta\mathrm{B}$ is the difference in $\mathrm{B}$ between the entangled pairs' locations.
The Bell mixing angle for the pair state \eqref{eqn:bell_state_t_evo} can be determined from a parity measurement involving the two-body spin correlator in the complementary basis given by
\begin{equation}
    \textrm{parity} = \expval{\hat{\sigma}_{x}^{(A)} \hat{\sigma}_{x}^{(B)}} = \cos 2\Phi,
    \label{eqn:parity_vs_corr}
\end{equation}
where $\hat{\sigma}_{x}^{(l)}$ is the $x$-basis Pauli matrix for the atom at location $l$ (see Fig.~\ref{fig:exp_schematic}).
Note that in our experiment we only directly measure $\hat{\sigma}_z$.
$\hat{\sigma}_x$ measurement is implemented by preceding the SG sequence ($\hat{\sigma}_z$) with a $\pi/2$-pulse, which maps the $\hat{\sigma}_x$-basis spin components into that of $\hat{\sigma}_z$ (see Fig.~\ref{fig:exp_schematic}(b)).

\begin{figure}[!htbp]
    \centering
    \includegraphics[width=8.6cm]{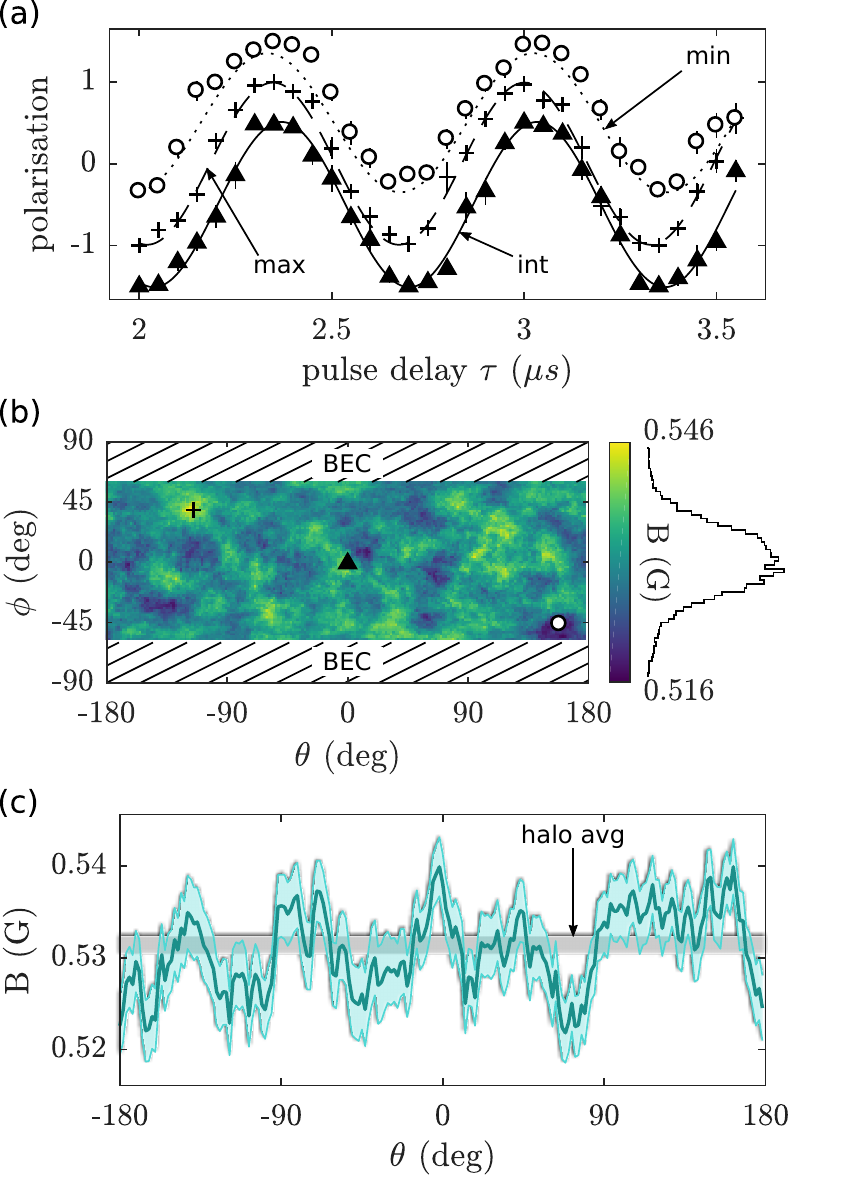}
    \caption{Experimental results for Ramsey-based 3D magnetic tomography.
    (a) Normalised polarisation $P$ (markers) and fitted Ramsey signal (lines) for selected regions on the halo indicated in (b).
    $\mathrm{B}(\mathbf{r})$ is determined by the fitted oscillation frequency.
    Data is offset vertically for clarity.
    (b) Spatial distribution of the measured magnetic field $\mathrm{B}$ on the scattering halo, and a histogram of $\mathrm{B}$.
    The hatched regions of the scattering halo indicate where the BECs saturate the detector and are ignored from data analysis.
    (c) Comparison between the spatially resolved (green line) estimation taken around equator ($\phi=0$) from (b), to the average field over the whole halo (grey band).
    All error bars and shaded regions indicate a $1\sigma$ standard error in the mean.}
    \label{fig:result_mag}
\end{figure}

\subsubsection{3D tomography of magnetic field}
Figure~\ref{fig:result_mag}(a) shows the observed Ramsey signal at representative locations on the scattering halo corresponding to where the maximum (max), minimum (min), and an intermediate (int) value magnetic fields were observed.
10 shots were taken at each interrogation time $\tau$, with a single-shot average of $\approx 640$ atoms scattered into an individual binning region, and $\approx 68000$ in the whole scattering halo, accounting for the detector efficiency.
We fit Eq.~\eqref{eqn:ramsey_p_vs_t} to the observed Ramsey signal around the halo (lines in Fig.~\ref{fig:result_mag}(a)), and reconstruct $\mathrm{B}(\mathbf{r})$ over the halo, as shown in Fig.~\ref{fig:result_mag}(b) using the equirectangular projection of the scattering halo.
The hatched areas in Fig.~\ref{fig:result_mag}(b) correspond to excluded regions of the scattering halo near the BECs ($|\phi| > 60^\circ$), which causes the detector to saturate.
The measured $\mathrm{B}(\mathbf{r})$ has no discernible spatial structure and is well described by a Gaussian distribution (shown to the right in Fig.~\ref{fig:result_mag}(b)) with a mean of $0.532$~G and a standard deviation of $\approx 4~\textrm{mG}$ over the interrogated region.
An independent measurement of the bias magnetic field, based on the Zeeman shift of the resonance of a two-photon Raman transition \cite{Moler1992}, yields $0.53(1)~\mathrm{G}$, in good agreement with this result.

Figure~\ref{fig:result_mag}(c) shows the reconstructed field strength around the equator ($\phi = 0$) of the halo (solid line) extracted from Fig.~\ref{fig:result_mag}(b), in comparison to the average over the entire halo (horizontal grey band).
The improvement in the measurement uncertainty $\Delta\mathrm{B}$ (width of shaded error bars) between the spatially resolved scheme of $3.4$~mG (integration half-cone angle $\alpha = 0.062\pi$) and completely spatially integrated measurement $0.9$~mG ($\alpha = \pi$) is due to the $1/\sqrt{N}$ shot-noise scaling attributed to the relative size of the integration volumes, since the field is effectively uniform.
Since $\Delta\mathrm{B}$ for each region is comparable to the variation seen across the halo (Gaussian width), this proof-of-principle result is unable to distinguish field inhomogeneity from the measurement noise.
Still however, this result serves to infer an estimate of the 3D spatial distribution of magnetic field gradient, albeit with rather crude accuracy, and thus forms an important calibration tool for the entanglement-based magnetic gradiometry scheme.
For completeness, we briefly note that the sensitivity can be vastly improved over the proof-of-principle result by either increasing the interrogation time $\tau$, the number of scattered atoms, and the total acquired experimental shots, or to adopt a more optimal parameter estimation strategy \cite{Ferrie2012}.

\subsubsection{Magnetic gradiometry with entangled atoms}
\begin{figure}[!htbp]
    \centering
    \includegraphics[width=8.6cm]{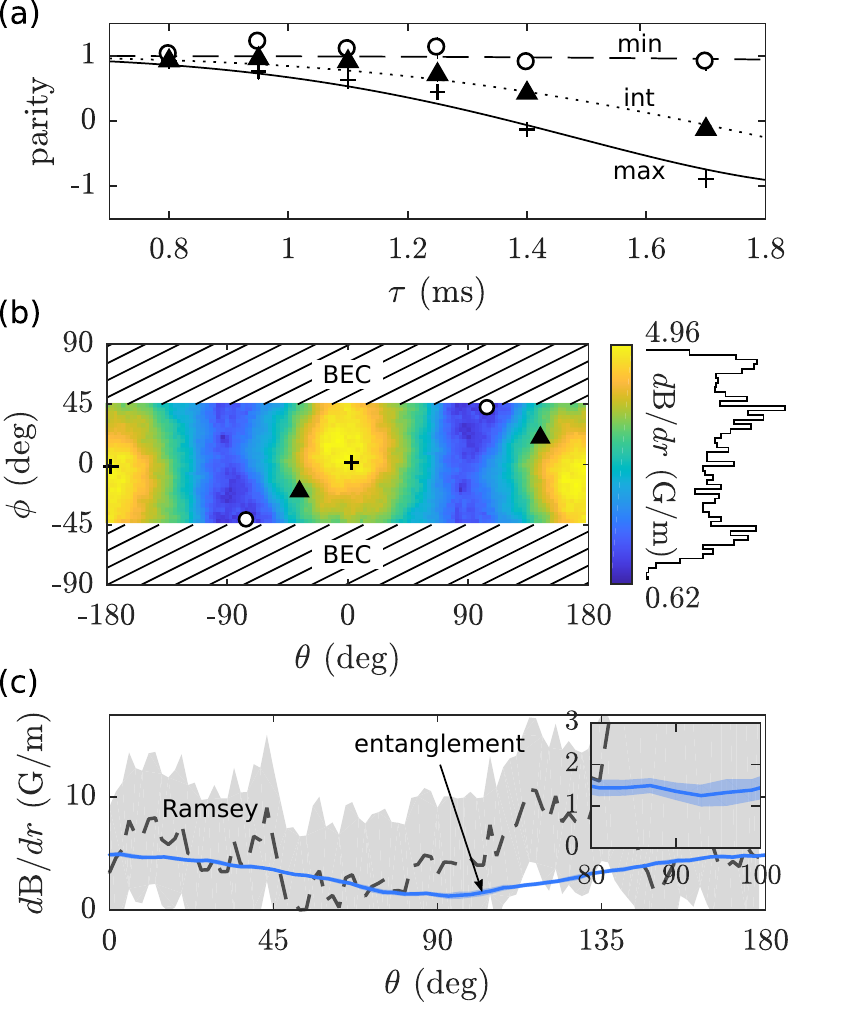}
    \caption{
    Experimental results for entanglement-based 3D magnetic gradiometry.
    (a) Time evolution of parity observed at selected antipodal regions on the halo [see markers in (b)] and fits from \eqref{eqn:Phi_dynamics} and \eqref{eqn:parity_vs_corr} (lines), giving the estimate of $d\mathrm{B}/dr$.
    (b) Spatial distribution of $d\mathrm{B}/dr$ on the scattering halo and its histogram (shown on right).
    The hatched regions contain atoms from BECs and are ignored in the data analysis.
    (c) Comparison of $d\mathrm{B}/dr$ between the entanglement-based scheme (blue line) and the Ramsey method (black dashed line), shown around the equator.
    The inset shows the graph zoomed-in for a comparison of uncertainties.
    Statistical uncertainties for the parity analysis were estimated by bootstrapping, and all error bars and shaded regions indicate a $1\sigma$ standard error in the mean.}
    \label{fig:result_grad}
\end{figure}

Figure~\ref{fig:result_grad}(a) shows the observed time-evolution of parity at three representative locations on the scattering halo, where the parity for $\Psi^\mp$ (anti-/symmetric) and superpositions take $\mp1$ and intermediate values, respectively (see \cite{SOMs} for the details of parity analysis).
The earliest time we have applied the $\pi/2$-pulse to the entangled halo is $0.8~\textrm{ms}$ after the collision sequence (see Fig.~\ref{fig:exp_schematic}(b)), which provides sufficient time for the BECs to fully separate ($t_{\textrm{sep}} = R_{\textrm{TF}}/2v_r \approx 0.4~\textrm{ms}$ \cite{SOMs}), ensuring no more pairs are scattered into the halo after the rotation.
The state of the pairs remain close to $\ket{\Psi^+}$ for short times after the collision ($\tau \lesssim 1~\mathrm{ms}$) regardless of their location on the halo (i.e. scattering angle), since there is insufficient spatial separation between the pairs, and thus relatively small asymmetry in the magnetic field.
At longer delays, we observe a strong scattering angle-dependent mixing of the Bell states, such that at $\tau=1.7~\textrm{ms}$ the halo simultaneously contains almost stationary regions of $\ket{\Psi^+}$ ($\circ$-marker), and regions of the orthogonal states $\ket{\Psi^-}$ ($+$-marker).

To explain this, we model $\mathrm{B}(\mathbf{r})$ up to linear terms around the centre of the halo ($D \lesssim 0.2~\textrm{mm}$), such that $\mathrm{B}(\mathbf{r}) = \mathrm{B}_0 + \nabla \mathrm{B} \cdot \mathbf{r}$.
Then, a pair $\ket{\Psi}$ counter-propagating at velocities $\pm\mathbf{v}_r$ evolves according to $d\Phi/dt = \gamma/2 \nabla\mathrm{B} \cdot (2\mathbf{v}_r t)$. 
In terms of $d\mathrm{B}/dr = \nabla\mathrm{B} \cdot \mathbf{v}_r/v_r$, the gradient of magnetic field strength along the scattering axis, gives
\begin{equation}
    \Phi(\tau) = \frac{\gamma v_r}{2} \frac{d\mathrm{B}}{dr}\tau^2.
    \label{eqn:Phi_dynamics}
\end{equation}

This model gives an excellent fit to the dynamics of the atom pairs (see Fig.~\ref{fig:result_grad}(a)), and explains the transition from almost stationary early dynamics from $\ket{\Psi^+}$ states to the diverging behaviour seen at later times, summarised by the quadratic dependency of $\Phi$ on $\tau$.
Figure~\ref{fig:result_grad}(b) shows the reconstructed spatial distribution of $d\mathrm{B}/dr$ based on the single free-parameter of the fitted model. 
The reconstructed spatial distribution in Fig.~\ref{fig:result_grad}(b) is consistent with our model of uniform gradient across the halo, which is estimated to be $\grad \mathrm{B} \approx 5.0(2) \hat{\mathbf{x}}~\textrm{mG/mm}$.
Indeed as expected from the uniform gradient model, we observe in Fig.~\ref{fig:result_grad}(b) a principal direction ($\approx x$-axis) along which $\mathrm{B}$-field gradient takes the maximum value, away from which the field becomes gradually more uniform, and regions of minimum gradient (albeit non-zero) is observed along the perpendicular directions on the scattering halo.
The equatorial distribution of $d\mathrm{B}/dr$ along with the measurement uncertainty is shown by the blue line in Fig.~\ref{fig:result_grad}(c), where the average uncertainty was $\Delta(d\mathrm{B}/dr) \approx 0.22$~mG/mm.

The entanglement-based result lies within $\approx 1\sigma$-error of the prediction based on our Ramsey method (see black dashed line in Fig.~\ref{fig:result_grad}(c)) evaluated with the same integration volume, which gives an uncertainty $\sqrt{2}\Delta\mathrm{B}/D \approx 6.2$~mG/mm ($\sqrt{2}$ factor arises from the rms sum of uncorrelated errors in estimating the differential).
A quantitative comparison of $d\mathrm{B}/dr$ between the two results, used other than as a test for consistency, such as comparing the sensitivities, is however quite complicated since resources such as $\tau$, $N$, and the spatial resolution need to be normalised, and are not important to the scope of this work, since we only use the Ramsey method as a calibration tool.

Since the SQL of Larmor precession-based magnetometers is given by $\Delta\mathrm{B} = 1/ \gamma \sqrt{N \tau}$ \cite{Budker2002}, the schemes demonstrated here (both low $N$ and $\tau$) do not compete favourably with state-of-the-art magnetometers in sensitivity.
A major technical limitation in our experiment is the inefficient interrogation duty-cycle of approximately $10^{-4}$ (entanglement) and $10^{-7}$ (Ramsey) in a single shot ($\approx 25$~s) which is mostly spent on creating a BEC.
In addition, there is a stringent requirement for a high efficiency detector in measuring the $N$-particle joint detection \cite{Jones2009}.
We find that by considering an imperfect $\eta$, the minimum phase uncertainties for the proposed Ramsey (coherent $N$ spin-1/2 state) and entanglement-based ($N=2$ Bell state) schemes are given by $1/\sqrt{N \eta}$ and $1/2\eta$, respectively \cite{SOMs}.
Therefore, the minimum phase uncertainty $\Delta\Phi$ achievable for our atom pairs in Bell states, and hence all derived quantities for applications, can surpass the SQL only when $\eta > 1/\sqrt{2}$ and achieve the Heisenberg-limit when $\eta=1$.
Rather, this investigation serves as a proof-of-concept demonstration of magnetometry using a unique matter-wave platform (entanglement in scattering halos) operating in the few particle regime.

To conclude, we have demonstrated two complementary quantum metrology schemes with an ultracold atomic scattering halo where the free-expansion dynamics of the ensemble was utilised for 3D tomography of the magnetic field and its gradient.
This marks the first application of the quantum correlation and entanglement in the scattering halo, from which we envisage extensions to quantum tests of general relativity \cite{Geiger2018}, quantum nonlocality with massive particles \cite{Lewis-Swan2015,Shin2018}, as well as in other schemes of quantum-enhanced metrology \cite{Wasak2018}.

\section*{Acknowledgements}
The authors would like to thank Michael Barson, Jan Chwedenczuk, Heyang Li, Samuel Nolan, and Tomasz Wasak for insightful discussions.
This work was supported through Australian Research Council (ARC) Discovery Project grants DP120101390, DP140101763 and DP160102337.
DKS and JAR are supported by an Australian Government Research Training Program Scholarship.
SSH is supported by ARC Discovery Early Career Researcher Award DE150100315.



%

\clearpage
\onecolumngrid
\appendix

\section{Supplemental Material}

\subsection{Velocity and spin resolved detection}
All the pulse sequences in our experiment are done within a few $\textrm{ms}$ after switching off the magnetic trap at $t = 0$, located $d\approx848~\textrm{mm}$ above the micro-channel plate - delay line detector (MCP-DLD) stack that resolves individual atoms' time $t^*$, and $x,y$ positions at impact, with corresponding resolutions of $\approx3~\mu\textrm{s}$, and $\approx120~\mu\textrm{m}$, respectively \cite{Henson2018}.
The 3D velocity $\mathbf{v}$ of atoms at $t \approx 0$ can be reconstructed by the free-fall relations $v_x = (x - x_0)/t^*$, $v_y = (y - y_0)/t^*$ for the $x,y$-components, where the 0-subscript refers to the variables at $t=0$.
The vertical component explicitly includes the acceleration due to gravity by $v_z = gt^*/2  - (d - z_0)/t^*$, where $z_0$ is measured relative to the centre of trap.
With typical trapping frequencies used in our experiment, the spatial extent of the source is less than $100~\mu\textrm{m}$, and therefore can be ignored (i.e. $x_0,~y_0,~z_0 \approx 0$) in comparison to the displacement due to the free-expansion [displacement from single-photon recoil $T^* \hbar k_\textrm{photon}/m \approx 38.3~\textrm{mm}$, where $m$ is the mass of a helium-4 atom, and $k_{\textrm{photon}} = 2\pi/1.083~\mathrm{\mu m}^{-1}$ is the wavevector of a Raman laser (see \cite{Shin2018})].
A Taylor's expansion of $v_z$ about an initially stationary state ($\mathbf{v}_0 \approx 0$) yields $v_z = gT_0\tau - \frac{d}{T_0}\tau^2 + O(\tau^3)$, where $\tau = (t^* - T^*)/T^*$ is the normalised relative time of arrival, and $T^* = \sqrt{2d/g} \approx 416~\textrm{ms}$ the time-of-flight of a stationary atom.
In our experiment, the collision geometry from the two-photon Raman process restricts $\tau$ to less than $0.03$, such that even the first order approximation $v_z \approx gT_0\tau$ is accurate to $1\%$. 

\begin{figure}[!htbp]
    \centering
    \includegraphics[width=6cm]{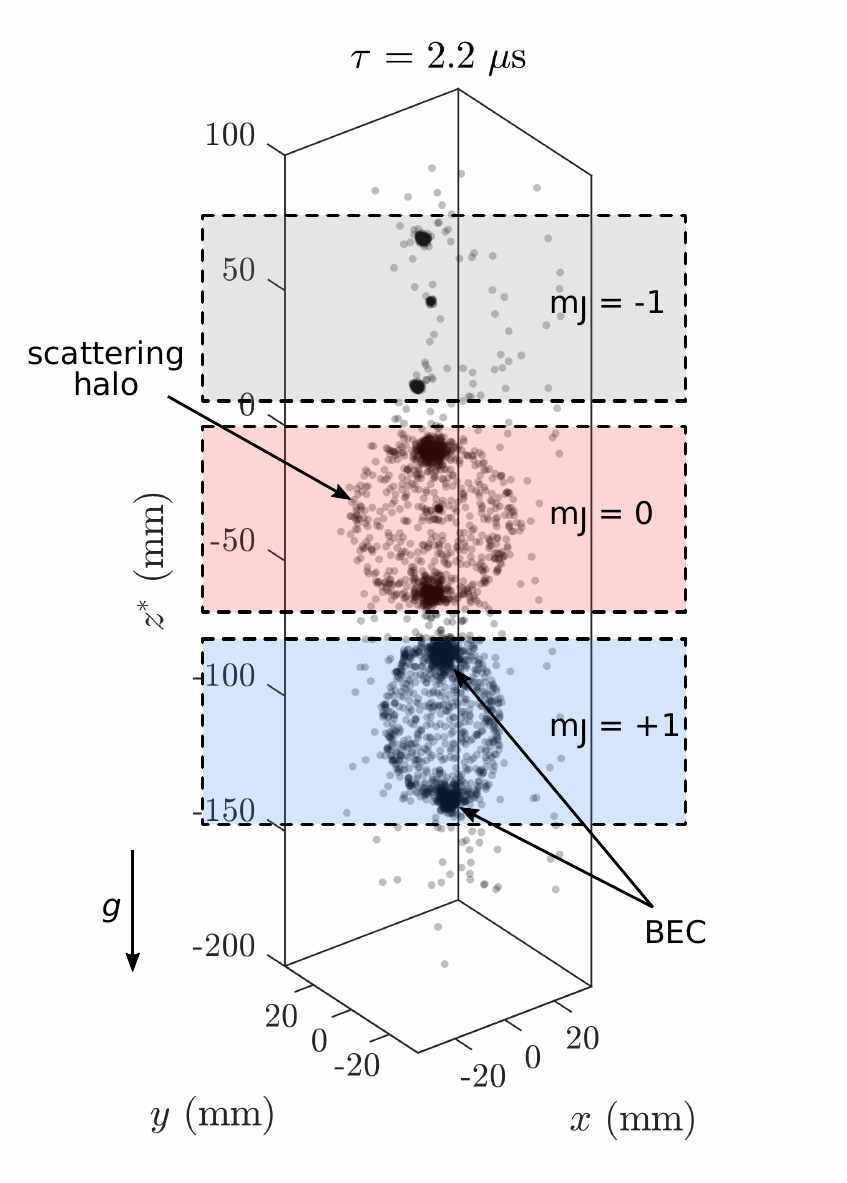}
    \caption{3D scatter plot from a single shot of the Ramsey interferometry experiment shown for $\tau = 2.2~\mathrm{\mu s}$.}
    \label{fig:zxy_3d}
\end{figure}

Figure~\ref{fig:zxy_3d} shows a typical reconstructed scatter plot of atoms from a single shot of our experiment, taken from the Ramsey interferometry experiment.
The time of arrival of each atom has been transformed to the relative vertical location by $z^* = v_z t^*$ for each atom. 
From the raw scatter data in Fig.~\ref{fig:zxy_3d}, we observe the collision BECs (six dense balls) and the scattering halos (faint spheres), distinguished by their $m_J$ states by the SG separation along $z$-axis.

\begin{figure}[!htbp]
    \centering
    \includegraphics[width=10cm]{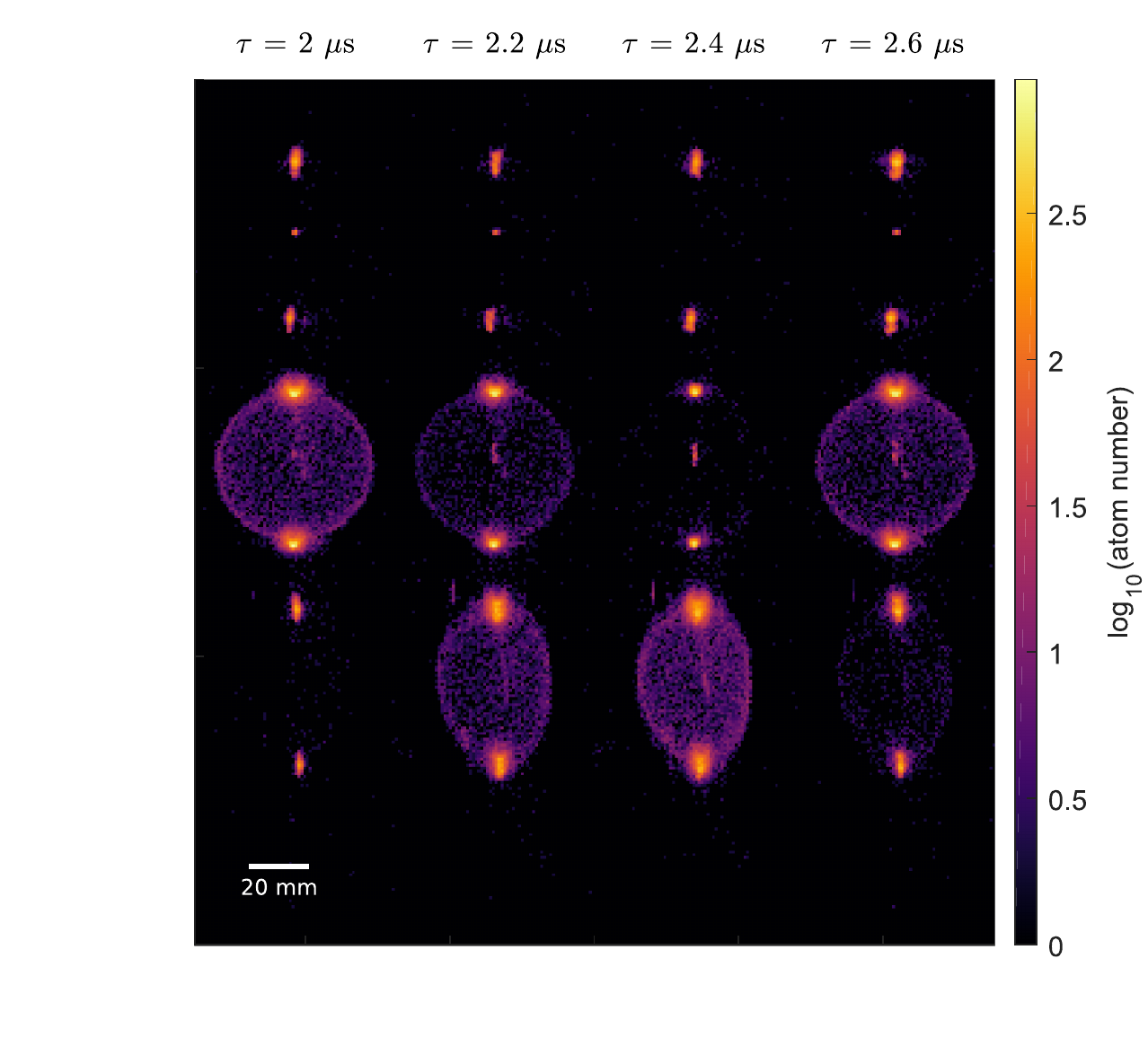}
    \caption{Single shot 2D density projection images in the $xz$-plane projection at various delays of the Ramsey interferometer.}
    \label{fig:2d_raw_ramsey}
\end{figure}

\subsubsection{Magnetic lensing}
As the interferometric phase in the Ramsey sequence is varied with pulse delay $\tau$ (see Fig.~\ref{fig:exp_schematic}(c) and Eq.~\eqref{eqn:ramsey_p_vs_t}), we observe the Ramsey fringe, namely the oscillation in the populations between the $m_J=0$ and $+1$ (see Fig.~\ref{fig:2d_raw_ramsey}).
The striking shape distortion of the $m_J=+1$ magnetically sensitive state compared to the $m_J=0$ state (zero magnetic moment) is due to a magnetic lensing effect (not to be confused with the Lorentz force on charged particles) from nonuniform stray magnetic fields in the vacuum chamber.
This effect was determined to occur much later in the atoms' trajectory with respect to the interferometry and SG sequence, and ultimately only complicates the velocity reconstruction procedure which is resolved as follows.

The simplest shape distortion by the magnetic lensing effect is to transform an initially spherical velocity distribution into an ellipsoid, predicted by the second-order inhomogeneity of magnetic field (i.e. the position dependence of $\grad \mathrm{B}$).
We thus reconstruct the original velocity of the atoms by first fitting the 3D density distribution of the scattering halo to an ellipsoid, and then transforming it to a unit sphere (radius normalised to $v_r$) by appropriately scaling the vector components along each fitted semi-axes.

\begin{figure}[!htbp]
    \centering
    \includegraphics[width=0.66\textwidth]{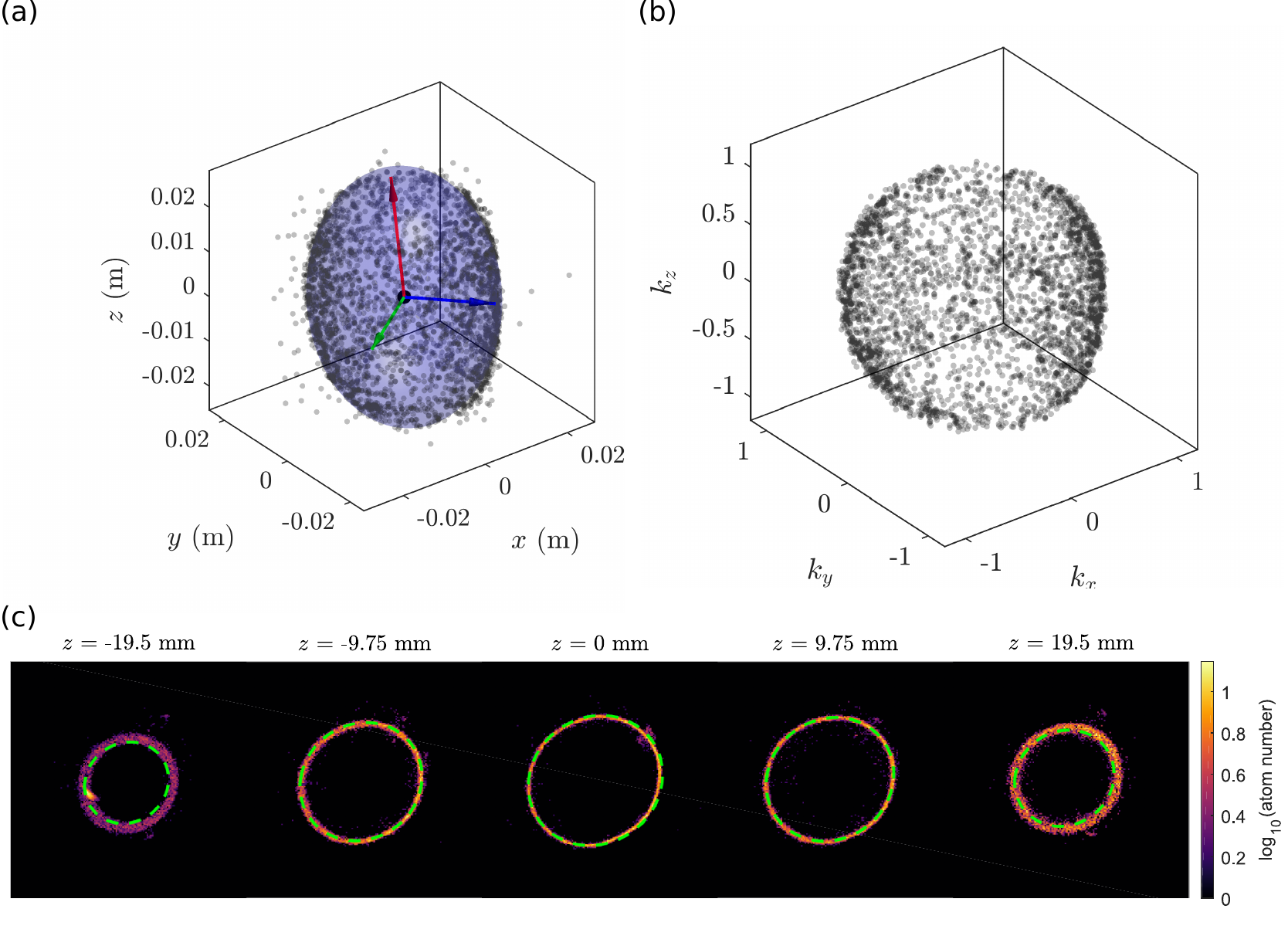}
    \caption{(a) Ellipsoid fit to the $m_J=+1$ scattering halo in the detector (position) space.
    (b) Post-processed normalised momentum distribution.
    (c) A comparison between the fitted ellipsoid (green dashed line) and the observed density profile of an $m_J=+1$ scattering halo, at various 2D slices taken perpendicular to the $z$-axis.}
    \label{fig:ellipfit_3d}
\end{figure}

Figure~\ref{fig:ellipfit_3d}(a) shows a subset of atoms (grey points) from the $m_J=+1$ scattering halo [truncated near the $\pm z$-poles and radially filtered to within $0.6 < r/r_0 < 1.2$ where $r_0$ is a first approximation of the halo's radius given by half the distance between the two collision BECs], and the fitted ellipsoid (blue surface), along with its centre (black point) and semi-axes (coloured arrows).
The ellipsoid gives an excellent fit to the $m_J=+1$ scattering halo as seen by the comparison between the 2D slices of density profiles along the $z$-axis and the fitted ellipsoid contours (see Fig.~\ref{fig:ellipfit_3d}(c)).
The post-processed momentum ($\mathbf{k}$-space) distribution of the distorted scattering halo is shown is Fig.~\ref{fig:ellipfit_3d}(b).

\begin{figure}[!htbp]
    \centering
    \includegraphics[width=7cm]{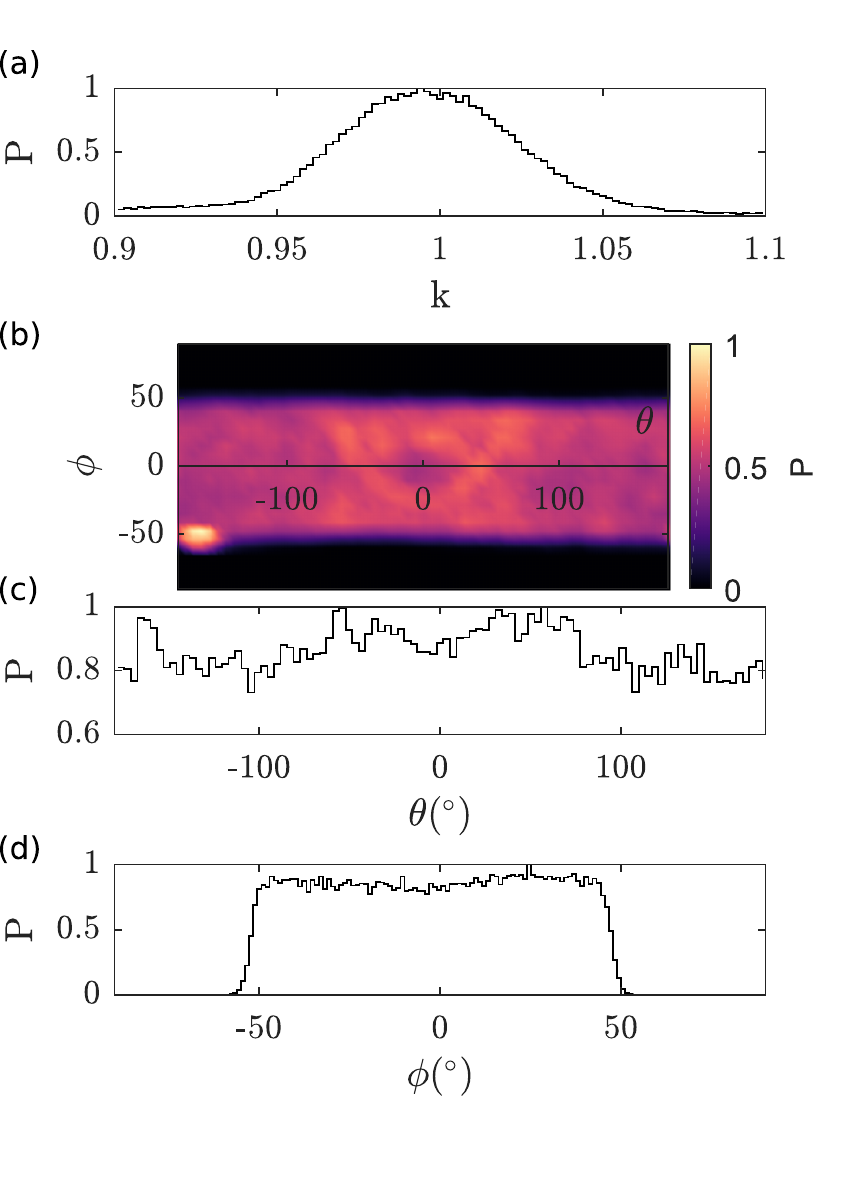}
    \caption{Characterisation of $\mathbf{k}$-space density distribution of the $m_J=+1$ scattering halo.
    (a) Radial distribution. $\mathrm{P}$ is the atom number density, normalised such that the maximum value is 1.
    (b) Spherical distribution. 
    The empty regions near the $\pm z$ poles ($\phi > 50^\circ$) are due to the truncation of the atoms near the BECs which is affected by detector saturation. 
    (c) Azimuthal distribution (elevation angle integrated).
    (d) Polar distribution (azimuthal angle integrated).}
    \label{fig:sph_dist_mJ1}
\end{figure}

Figure~\ref{fig:sph_dist_mJ1} shows the $\mathbf{k}$-space density distribution of the $m_J=+1$ scattering halo as shown in Fig.~\ref{fig:ellipfit_3d}(b). 
The $m_J = +1$ scattering halo is indeed a thin spherical shell in $\mathbf{k}$-space, identical to the $m_J = 0$ halo, with a normalised rms width $dk \approx 0.03$ (see Fig.~\ref{fig:sph_dist_mJ1}(a)) corresponding to the momentum width of the colliding BECs \cite{Oegren2009}, and is uniformly distributed around the sphere (see Fig.~\ref{fig:sph_dist_mJ1}(b--d)).

\subsection{Entanglement-based metrology}

\subsubsection{Characterisation of pairwise correlations}
\begin{figure}[!htbp]
    \centering
    \includegraphics[width=0.8\textwidth]{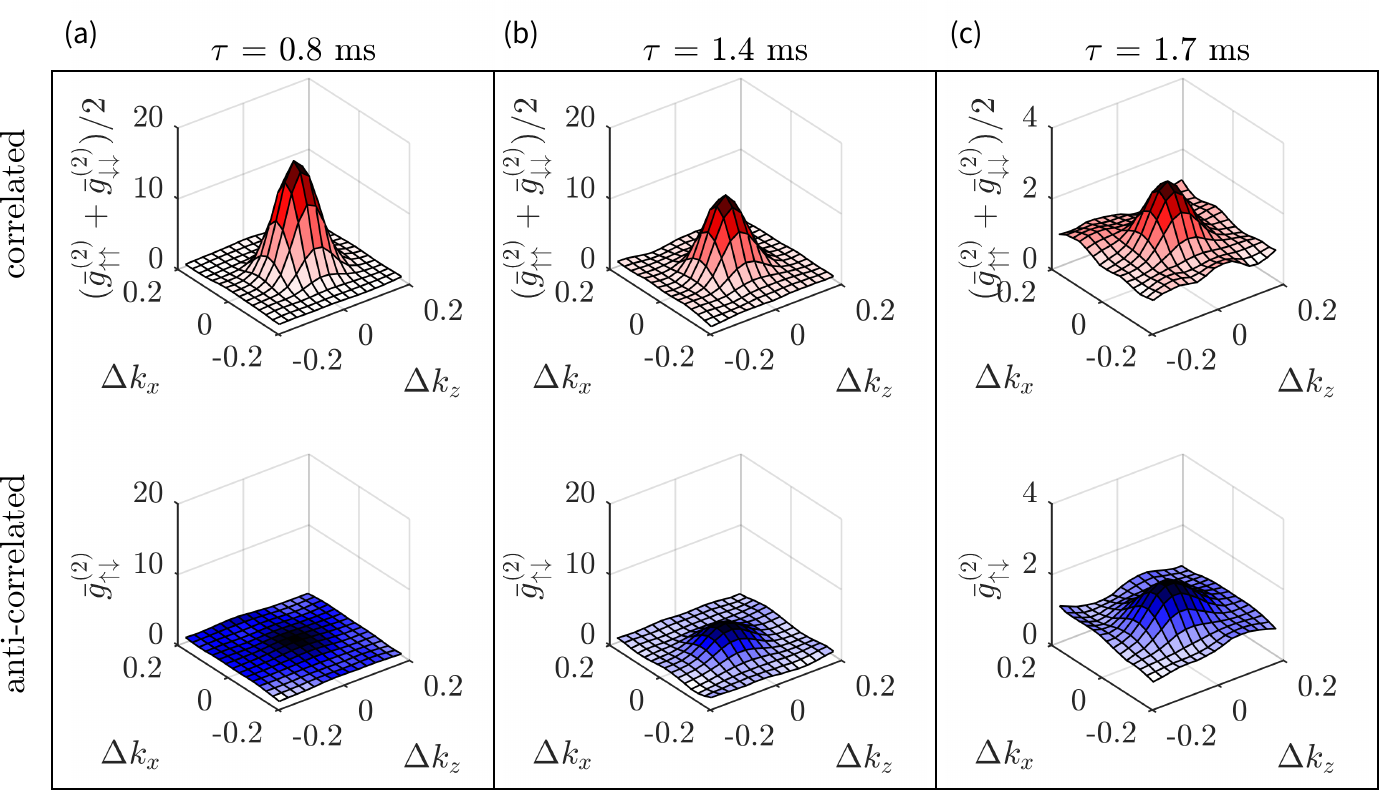}
    \caption{Integrated second order correlation functions across back-to-back momentum pairs in the scattering halo.
    Spin correlated (top), and anti-correlated $\bar{g}^{(2)}$ correlation functions (bottom) evaluated for (a) $\tau=0.8$, (b) $1.4$, and (c) $1.7$~ms after the collision pulse.
    The 2D profiles are taken in the $zx$-plane (in $\mathbf{k}$-space) where $\Delta k_y \approx 0$.
    }
    \label{fig:g2_integrated}
\end{figure}

To characterise the parity of the spin entangled scattering halo we rely on the two-particle correlation functions between atoms on opposite sides of the halo with either parallel or anti-parallel spin-pairing, given by
\begin{equation}
\bar{g}^{(2)}_{
ij} (\Delta \mathbf{k}) = \frac{\sum\nolimits_{\mathbf{k} \in V} \left\langle \hat{n}_{\mathbf{k},i} \hat{n}_{-\mathbf{k} + \Delta\mathbf{k},j} \right\rangle_{\pi/2}}{\sum\nolimits_{\mathbf{k} \in V}  \left\langle \hat{n}_{\mathbf{k},i} \right\rangle_{\pi/2} \left\langle \hat{n}_{-\mathbf{k} + \Delta\mathbf{k},j} \right\rangle_{\pi/2}},
\label{eqn:g2}
\end{equation}
where $\pi/2$ subscript denotes that the expectation value is taken for the $\pi/2$ rotated state (i.e. spin correlation in $\hat{\sigma}_x$-basis), $i, j\in \{\uparrow,\downarrow\}$ the spin states, $\hat{n}_{\mathbf{q}, m}$ the number of atoms with momentum $\mathbf{q}$ and spin $m$, and $V$ the integration volume in momentum space corresponding to the truncated scattering halo.
Note that the bar in the definition of the second order correlation function is to signify that the correlator has been integrated for all pairs in the scattering halo.
Figure~\ref{fig:g2_integrated} compares this set of second order correlation functions for the scattering halo created by a collision of $\ket{\uparrow}$ and $\ket{\downarrow}$ BECs ($\hat{\sigma}_z$-basis), interrogated between $\tau = 0.8~\textrm{ms}$ and $1.7~\textrm{ms}$ after the collision is initiated (see main text for details).
We summarise the set of $\bar{g}^{(2)}$ functions by (1) the correlated configurations $\uparrow\uparrow$ and $\downarrow\downarrow$ by their average $(\bar{g}^{(2)}_{\uparrow\uparrow} + \bar{g}^{(2)}_{\downarrow\downarrow})/2$ (see top row of Fig.~\ref{fig:g2_integrated}), and (2) the anti-correlated configuration $\bar{g}^{(2)}_{\uparrow\downarrow}$ (see bottom row of Fig.~\ref{fig:g2_integrated}). 
The large signal for correlated pairs in spin at $\tau = 0.8~\textrm{ms}$ (see Fig.~\ref{fig:g2_integrated}(a)) relative to anti-correlated events (note that $\bar{g}^{(2)} = 1$ for uncorrelated pair events), is expected at early times after the creation of the $\ket{\Psi^+}$-halo (see main text).
In addition, at longer $\tau$ we observe a gradual increase in the relative observation of anti-correlated pairs (see Figs.~\ref{fig:g2_integrated}(b,c)), such that by $\tau = 1.7~\textrm{ms}$ the $\bar{g}^{(2)}$ amplitudes between different spin configurations become comparable in magnitude. 
Note that the difference in the magnitude of correlation amplitudes measured at different $\tau$ is due to the difference in the average number of atoms in the scattering halo between different experiments, which has an inverse relationship with the $\bar{g}^{(2)}$ amplitude \cite{Hodgman2017}. 
We show in the following section that the determination of Bell mixing angles requires only the knowledge of ratios of $\bar{g}^{(2)}$ between different spin configurations, and therefore insensitive to the absolute scattered atom number.

\begin{figure}[!htbp]
    \centering
    \includegraphics[width=\textwidth]{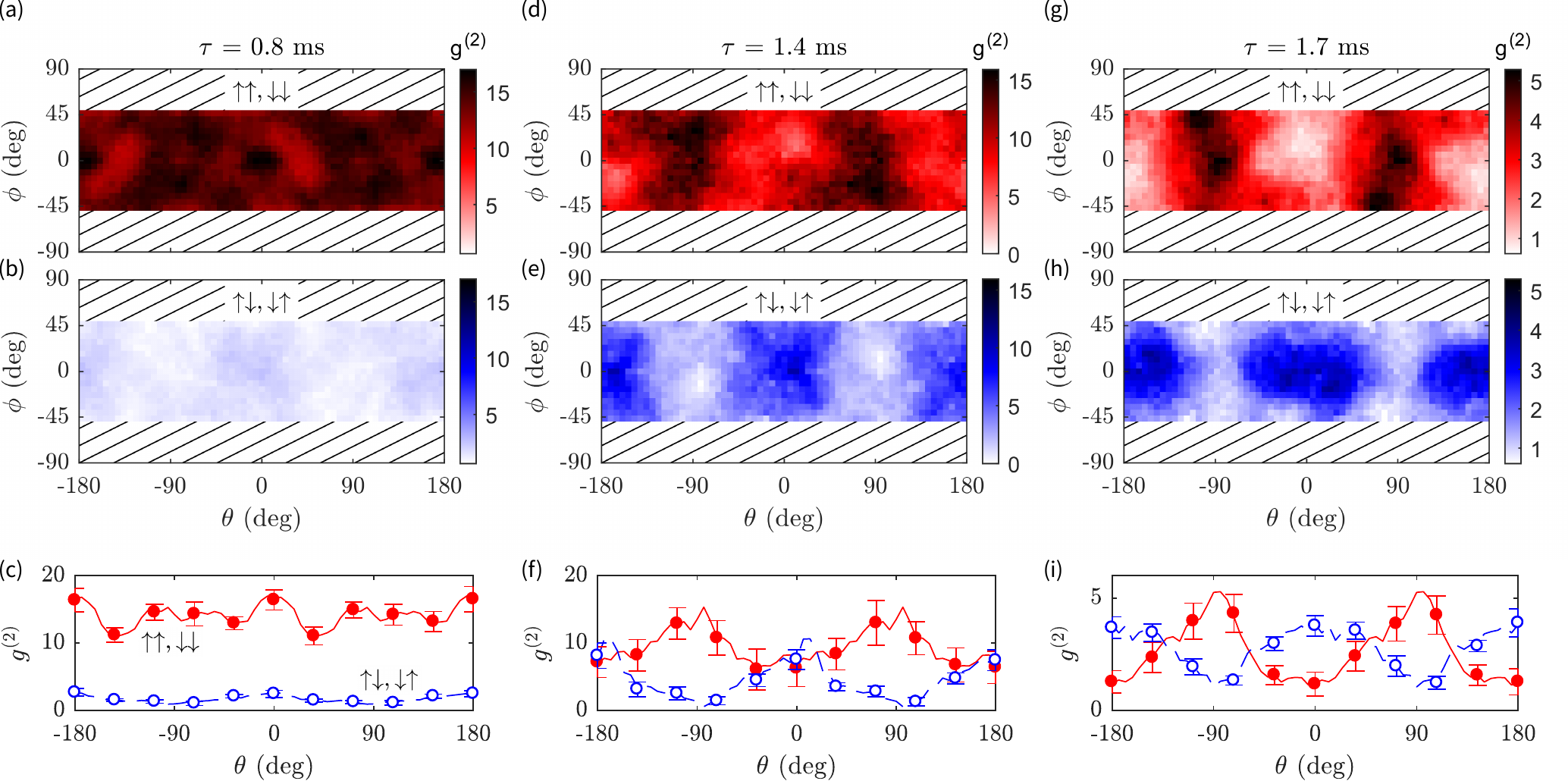}
    \caption{Localised second order correlation function across back-to-back momentum pairs in the scattering halo.
    (a) Spin correlated configuration.
    (b) Spin anti-correlated configuration.
    (c) Comparison of $g^{(2)}$ around the equator ($\phi=0$) for correlated (red solid line) and anti-correlated configurations (blue dashed line).
    (a--c) are evaluated $\tau = 0.8~\textrm{ms}$ after the collision pulse.
    (d--f) and (g--i) are similar to (a--c), for $\tau = 1.4$ and $1.7~\textrm{ms}$, respectively.
    Error bars indicate a $1\sigma$ standard error in the mean estimated from bootstrapping.
    }
    \label{fig:g2_localised}
\end{figure}

In order to determine the scattering angle or spatially resolved pairwise correlations, and thus magnetic field gradients, the integration volume chosen in \eqref{eqn:g2} must be localised around a specific region in the scattering halo.
We introduce the localised two-particle correlation function across back-to-back pairs in $\mathbf{k}$-space
\begin{equation}
    g_{ij}^{(2)}(\mathbf{q}) = \frac{\sum\nolimits_{\mathbf{k} \in \delta V} \left\langle \hat{n}_{\mathbf{k},i} \hat{n}_{-\mathbf{k},j} \right\rangle_{\pi/2} }{\sum\nolimits_{\mathbf{k} \in \delta V}  \left\langle \hat{n}_{\mathbf{k},i} \right\rangle_{\pi/2} \left\langle \hat{n}_{-\mathbf{k},j} \right\rangle_{\pi/2}},
\label{eqn:g2_loc}
\end{equation}
as an adequate extension to \eqref{eqn:g2} where the localised integration volume $\delta V$ around $\mathbf{q}$ is as follows.
We used a double-cone oriented along $\pm\mathbf{q}$ as $\delta V$, with half-cone angle $\alpha = \pi/10$ (chosen so that the bin size at interrogation is equal to the spatial resolution).
Figure~\ref{fig:g2_localised} shows the distribution of $g^{(2)}$ measured for various $\tau$, where we clearly observe the scattering angle dependent time-evolution.
Figures~\ref{fig:g2_localised}(a-c) reveal that shortly after their creation ($\tau=0.8~\textrm{ms}$), atom pairs are mostly spin-correlated in the $\hat{\sigma}_x$-basis, independent of their scattering angle, verifying our claim that all pairs are approximately $\ket{\Psi^+}$. 
This result was exploited in \cite{Shin2018} so that the entire entangled scattering halo could be used for parallel realisations of a Bell test.
At longer evolution times we observe a strong scattering angle dependent mixing between $\ket{\Psi^{\pm}}$ states (see Figs.~\ref{fig:g2_localised}(d-i)), such that by $\tau = 1.7~\textrm{ms}$ distinct regions are occupied almost purely by either $\ket{\Psi^+}$ or $\ket{\Psi^-}$ (see Fig.~\ref{fig:g2_localised}(i)).

\subsubsection{Parity analysis}
From \eqref{eqn:parity_vs_corr} the parity of a single two-qubit state in a superposition of $\ket{\Psi^+}$ and $\ket{\Psi^-}$ can be determined based on the correlation of the pair's spins in the complementary basis such as $\hat{\sigma}_x$.
In this section, we explain the method used to evaluate the scattered pairs' parity which even accounts for multiple-pair mode occupation.

In the general case for arbitrary number of pairs scattered across back-to-back locations on the halo, we extend the idealised expression \eqref{eqn:parity_vs_corr} by extending the spin-1/2 operators $\hat{\sigma}_l$ to the generalised angular momentum operators $\hat{J}_l$ following \cite{Wasak2018}.
Spin resolved atom number measurements, based on the SG readout with the Raman rotation to select the basis of measurement, gives the angular momentum components
\begin{equation}
    J_l{(\mathbf{q})} = \left( n_{l,\uparrow}{(\mathbf{q})} - n_{l,\downarrow}{(\mathbf{q})} \right) /2,
\end{equation}
and the total atom number
\begin{equation}
    N{(\mathbf{q})} = \left( n_{l,\uparrow}{(\mathbf{q})} + n_{l,\downarrow}{(\mathbf{q})} \right) /2,
\end{equation}
such that the $x$-basis correlation coefficient for atoms scattered along $\pm\mathbf{k}$ is
\begin{equation}
    E_{xx}(\mathbf{k}) = \frac{\expval{J_x{(\mathbf{k})} J_x{(-\mathbf{k})}}}{\expval{N{(\mathbf{k})} N{(-\mathbf{k})}}},
    \label{eqn:general_corr}
\end{equation}
where $n_{l,i}{(\mathbf{q})}$ is the number of atoms in momentum mode $\mathbf{q}$, with spin $i$ in $\hat{\sigma}_l$-basis (see \cite{Wasak2018} for details).

It can be shown that \eqref{eqn:general_corr} can be written in terms of the localised correlation functions $g^{(2)}_{ij}$ as
\begin{equation}
    E_{xx} = \frac{g^{(2)}_{\uparrow\uparrow} - g^{(2)}_{\uparrow\downarrow} - g^{(2)}_{\downarrow\uparrow} + g^{(2)}_{\downarrow\downarrow}}
        {g^{(2)}_{\uparrow\uparrow} + g^{(2)}_{\uparrow\downarrow} + g^{(2)}_{\downarrow\uparrow} + g^{(2)}_{\downarrow\downarrow}},
    \label{corr_g2}
\end{equation}
where the $g^{(2)}$'s are implied to be in $\hat{\sigma}_x$-basis and localised at $\mathbf{k}$ (see \cite{Shin2018} for the derivation).

In the case of a spontaneous pair source, such as the scattering halo, the denominator of \eqref{corr_g2} is the spin integrated correlation amplitude (insensitive to Bell mixing angle), which is inversely proportional to the average atom number in a scattering mode $\bar{n}$ \cite{Perrin2008}:
\begin{equation}
    G = \left( g^{(2)}_{\uparrow\uparrow} + g^{(2)}_{\uparrow\downarrow} + g^{(2)}_{\downarrow\uparrow} + g^{(2)}_{\downarrow\downarrow} \right)/4 = 1 + 1/2\bar{n}.
    \label{g2_vs_n}
\end{equation}
On the other hand, the numerator is dependent on the pairs' mixing angle \eqref{eqn:bell_state_t_evo} by
\begin{equation}
    \left( g^{(2)}_{\uparrow\uparrow} - g^{(2)}_{\uparrow\downarrow} - g^{(2)}_{\downarrow\uparrow} + g^{(2)}_{\downarrow\downarrow} \right)/4 = \frac{\cos 2\Phi}{2\bar{n}}, 
\end{equation}
which is a parity-like signal \eqref{eqn:parity_vs_corr} with amplitude $1/2\bar{n}$, and the extrema correspond to anti-/symmetric states $\ket{\Psi^\mp}$, respectively, such that
\begin{equation}
    E_{xx} =  \frac{1}{1 + 2\bar{n}} \cos2\Phi.
\end{equation}

The effect of non-ideal purity of the pair source on $E_{xx}$ is therefore a reduction by a factor $(1 - 1/G)$ \eqref{g2_vs_n} with respect to the ideal case for a single pair \eqref{eqn:parity_vs_corr} ($G \gg 1$), predicted by the Bogoliubov scattering theory of colliding BECs \cite{Wasak2018}.
This result can be used to scale the measured correlation coefficient to estimate the parity of a single pair as is shown in Fig.~\ref{fig:result_grad}(a) by
\begin{equation}
    \mathrm{parity} = (1-1/G)^{-1} E_{xx},
\end{equation}
and allows faster data acquisition by using a brighter pair source.

\subsection{Spatial resolution from free-expansion of scattering halo} 
Here we estimate the spatial resolution achievable by atom interferometry schemes based on a freely-expanding scattering halo.
For simplicity, our analysis is in one dimension in space and gravity is ignored.

Suppose the scattering halo is created at $t=0$ from a collision of two BECs, such that a scattered atom's initial spatial location $r$ and velocity $v$ are normally distributed by
$r\sim\mathcal{N}(\mu=0,\sigma^2)$, and $v\sim\mathcal{N}(\mu=v_0,\sigma_v^2)$.
Such probability distribution of the initial position and velocity can be determined from properties of the BECs.
For simplicity we take the Thomas-Fermi approximation for the colliding BECs such that $\sigma \approx R_\textrm{TF}/\sqrt{2}$ (the factor of $1/\sqrt{2}$ is due to the density-squared scaling for two-body collision rates) and $\sigma_v \approx 2\hbar/mR_\textrm{TF}$ \cite{Oegren2009}, where $R_\textrm{TF}$ is the Thomas-Fermi radius, and $m$ is mass of the atom. 
Based on the parameters for our experiment, where the condensate number is $N \sim 10^5$, and trapping frequency $\bar\omega \approx 2\pi \cdot 20~\textrm{Hz}$, we obtain $R_\textrm{TF} \approx 50~\mathrm{\mu m}$ and $\sigma_v \approx 0.6~\textrm{mm/s}$.

Assuming that the scattered atoms do not interact strongly with the BECs and with each other, they propagate as free particles whereby an atom's position in the scattering halo after an arbitrary evolution time $t$ is given by $s(t) = r + vt$, thus $s\sim\mathcal{N}(\mu=v_0 t, \sigma^2 + \sigma_v^2 t^2)$.
With an ideal detection scheme at $t=T$, every atom's position in the scattering halo at this time $s(T) = S$ can be measured, whereas the magnetic field interrogation occurs at a previous time $t^*$.

The spatial resolution of such scheme corresponds to the spatial uncertainty of atom's positions at the interrogation time $s(t^*) = s^*$, based on all possible trajectories to the particular measurement outcome.
Thus, the conditional probability distribution of the atom's location at $t^*$ is given by
\begin{equation}
    \begin{split}
    P(s^*|S) &= P(s(t^*) = s^*) | s(T) = S) \\
            &=\frac{P( (s(t^*) = s^*) \cap (s(T) = S) )}{P(s(T) = S)}.
    \end{split}
    \label{eqn:int_spatial_prob_dist}
\end{equation}
The joint event probability (numerator) corresponds to a unique trajectory, with the initial position $r' = s^* - t^*v'$ and velocity $v' = (S-s^*)/(T-t^*)$, such that 
\begin{equation}
\begin{split}
        P( s^* \cap S ) &= P(r = r' \cap v = v')    \\
        &= P(r=r') P(v=v').
\end{split}
\label{eqn:trajectory}
\end{equation}
Substituting \eqref{eqn:trajectory} into \eqref{eqn:int_spatial_prob_dist} and evaluating the probability function for $s^*$ gives a normal distribution with parameters
\begin{subequations}
\begin{align}
    \expval{s^*}  &= \frac{(\tau + \xi^2)S - \sigma\xi w^{-1}(1-\tau)}{1 + \xi^2}  \\
    \Delta s^* &= \frac{\sigma (1 - \tau)}{\sqrt{1 + \xi^2}},
\end{align}
\label{eqn:spat_dist_params}
\end{subequations}
where $\tau = t^* / T$, $\xi = \sigma / \sigma_vT$ and $w = \sigma_v / v_0$ are the dimensionless parameters characterising the scheme.

In the far-field detection regime ($\xi \ll 1$) and with an ultracold scattering halo ($w \ll 1$), the mean position and spatial uncertainty at the interrogation time simultaneously simplify to first order
\begin{subequations}
\begin{align}
    \expval{s^*}_\infty &= \tau S  \\
    \Delta s^*_\infty &= \sigma (1 - \tau).
\end{align}
\label{eqn:spat_dist_params_ffhalolim}
\end{subequations}
Our experimental parameters are $\tau \approx 0.01$, $\xi \approx 0.1$, and $w \approx 0.03$, so that we can assume the evolution of the scattering halo to be a uniform expansion in time, where the uncertainty is limited by the size of the BEC \eqref{eqn:spat_dist_params_ffhalolim}.

\end{document}